\documentclass[12pt,a4paper]{article}

\usepackage{ifthen} 
\newboolean{pdflatex}
\setboolean{pdflatex}{true} 

\newboolean{articletitles}
\setboolean{articletitles}{true} 

\newboolean{uprightparticles}
\setboolean{uprightparticles}{false} 

\newboolean{inbibliography}
\setboolean{inbibliography}{false} 

\usepackage{microtype}
\usepackage{lineno}  
\usepackage{xspace} 

\usepackage{graphicx}  
\usepackage{color}
\usepackage{colortbl}
\graphicspath{{./figs/}} 

\usepackage{amsmath} 
\usepackage{amssymb}
\usepackage{amsfonts}
\usepackage{upgreek} 

\newcommand*\patchAmsMathEnvironmentForLineno[1]{%
\expandafter\let\csname old#1\expandafter\endcsname\csname #1\endcsname
\expandafter\let\csname oldend#1\expandafter\endcsname\csname
end#1\endcsname
 \renewenvironment{#1}%
   {\linenomath\csname old#1\endcsname}%
   {\csname oldend#1\endcsname\endlinenomath}%
}
\newcommand*\patchBothAmsMathEnvironmentsForLineno[1]{%
  \patchAmsMathEnvironmentForLineno{#1}%
  \patchAmsMathEnvironmentForLineno{#1*}%
}
\AtBeginDocument{%
\patchBothAmsMathEnvironmentsForLineno{equation}%
\patchBothAmsMathEnvironmentsForLineno{align}%
\patchBothAmsMathEnvironmentsForLineno{flalign}%
\patchBothAmsMathEnvironmentsForLineno{alignat}%
\patchBothAmsMathEnvironmentsForLineno{gather}%
\patchBothAmsMathEnvironmentsForLineno{multline}%
}

\usepackage{hyperref}    
\usepackage[all]{hypcap} 

\def\lhcb {\mbox{LHCb}\xspace}
\def\ux85 {\mbox{UX85}\xspace}

\ifthenelse{\boolean{uprightparticles}}%
{

 \def\Pchi        {\ensuremath{\upchi}\xspace}                 
 \def\Ppsi        {\ensuremath{\uppsi}\xspace}

 \def\PDelta      {\ensuremath{\Delta}\xspace}                 
 \def\PXi      {\ensuremath{\Xi}\xspace}                 
 \def\PLambda      {\ensuremath{\Lambda}\xspace}                 
 \def\PSigma      {\ensuremath{\Sigma}\xspace}                 
 \def\POmega      {\ensuremath{\Omega}\xspace}                 
 \def\PUpsilon      {\ensuremath{\Upsilon}\xspace}                 
 
 \def\PB      {\ensuremath{\mathrm{B}}\xspace}                 
                  
 \def\PD      {\ensuremath{\mathrm{D}}\xspace}

 \def\PJ      {\ensuremath{\mathrm{J}}\xspace}                 
 \def\PK      {\ensuremath{\mathrm{K}}\xspace}

 \def\Pb      {\ensuremath{\mathrm{b}}\xspace}                 
 \def\Pc      {\ensuremath{\mathrm{c}}\xspace}

 \def\Pi      {\ensuremath{\mathrm{i}}\xspace}

}
{

 \def\Pchi        {\ensuremath{\chi}\xspace}                 
 \def\Ppsi        {\ensuremath{\psi}\xspace}                 
                  
 \mathchardef\PDelta="7101
 \mathchardef\PXi="7104
 \mathchardef\PLambda="7103
 \mathchardef\PSigma="7106
 \mathchardef\POmega="710A
 \mathchardef\PUpsilon="7107
                  
 \def\PB      {\ensuremath{B}\xspace}                 
                  
 \def\PD      {\ensuremath{D}\xspace}

 \def\PJ      {\ensuremath{J}\xspace}                 
 \def\PK      {\ensuremath{K}\xspace}

 \def\Pb      {\ensuremath{b}\xspace}                 
 \def\Pc      {\ensuremath{c}\xspace}

 \def\Pi      {\ensuremath{i}\xspace}

}

\def\cquark    {\ensuremath{\Pc}\xspace}

\def\bquark    {\ensuremath{\Pb}\xspace}

\def\kaon  {\ensuremath{\PK}\xspace}
  \def\Kbar  {\kern 0.2em\overline{\kern -0.2em \PK}{}\xspace}

\def\Kz    {\ensuremath{\kaon^0}\xspace}
\def\Kzb   {\ensuremath{\Kbar^0}\xspace}
\def\KzKzb {\ensuremath{\Kz \kern -0.16em \Kzb}\xspace}
\def\Kp    {\ensuremath{\kaon^+}\xspace}
\def\Km    {\ensuremath{\kaon^-}\xspace}

\def\KpKm  {\ensuremath{\Kp \kern -0.16em \Km}\xspace}

  \def\Dbar    {\kern 0.2em\overline{\kern -0.2em \PD}{}\xspace}
\def\D       {\ensuremath{\PD}\xspace}

\def\Dz      {\ensuremath{\D^0}\xspace}
\def\Dzb     {\ensuremath{\Dbar^0}\xspace}
\def\DzDzb   {\ensuremath{\Dz {\kern -0.16em \Dzb}}\xspace}
\def\Dp      {\ensuremath{\D^+}\xspace}
\def\Dm      {\ensuremath{\D^-}\xspace}

\def\DpDm    {\ensuremath{\Dp {\kern -0.16em \Dm}}\xspace}

  \def\Bbar    {\kern 0.18em\overline{\kern -0.18em \PB}{}\xspace}

\def\jpsi     {\ensuremath{{\PJ\mskip -3mu/\mskip -2mu\Ppsi\mskip 2mu}}\xspace}
\def\psitwos  {\ensuremath{\Ppsi{(2S)}}\xspace}

\def\chiczero {\ensuremath{\Pchi_{\cquark 0}}\xspace}
\def\chicone  {\ensuremath{\Pchi_{\cquark 1}}\xspace}
\def\chictwo  {\ensuremath{\Pchi_{\cquark 2}}\xspace}
  \def\Y#1S{\ensuremath{\PUpsilon{(#1S)}}\xspace}

\def\Lbar {\ensuremath{\kern 0.1em\overline{\kern -0.1em\PLambda}}\xspace}

\def\AT#1     {\ensuremath{A_{\mathrm{T}}^{#1}}\xspace}           

\def\C#1      {\ensuremath{\mathcal{C}_{#1}}\xspace}                       
\def\Cp#1     {\ensuremath{\mathcal{C}_{#1}^{'}}\xspace}                    
\def\Ceff#1   {\ensuremath{\mathcal{C}_{#1}^{\mathrm{(eff)}}}\xspace}        
\def\Cpeff#1  {\ensuremath{\mathcal{C}_{#1}^{'\mathrm{(eff)}}}\xspace}       
\def\Ope#1    {\ensuremath{\mathcal{O}_{#1}}\xspace}                       
\def\Opep#1   {\ensuremath{\mathcal{O}_{#1}^{'}}\xspace}                    

\newcommand{\tev}{\ensuremath{\mathrm{\,Te\kern -0.1em V}}\xspace}
\newcommand{\gev}{\ensuremath{\mathrm{\,Ge\kern -0.1em V}}\xspace}
\newcommand{\mev}{\ensuremath{\mathrm{\,Me\kern -0.1em V}}\xspace}
\newcommand{\kev}{\ensuremath{\mathrm{\,ke\kern -0.1em V}}\xspace}
\newcommand{\ev}{\ensuremath{\mathrm{\,e\kern -0.1em V}}\xspace}
\newcommand{\gevc}{\ensuremath{{\mathrm{\,Ge\kern -0.1em V\!/}c}}\xspace}
\newcommand{\mevc}{\ensuremath{{\mathrm{\,Me\kern -0.1em V\!/}c}}\xspace}
\newcommand{\gevcc}{\ensuremath{{\mathrm{\,Ge\kern -0.1em V\!/}c^2}}\xspace}
\newcommand{\gevgevcccc}{\ensuremath{{\mathrm{\,Ge\kern -0.1em V^2\!/}c^4}}\xspace}
\newcommand{\mevcc}{\ensuremath{{\mathrm{\,Me\kern -0.1em V\!/}c^2}}\xspace}

\def\mum  {\ensuremath{\,\upmu\rm m}\xspace}

\def\nb {\ensuremath{\rm \,nb}\xspace}

\def\pb {\ensuremath{\rm \,pb}\xspace}
\def\invpb {\ensuremath{\mbox{\,pb}^{-1}}\xspace}

\def\invfb   {\ensuremath{\mbox{\,fb}^{-1}}\xspace}

\def\gsim{{~\raise.15em\hbox{$>$}\kern-.85em
          \lower.35em\hbox{$\sim$}~}\xspace}
\def\lsim{{~\raise.15em\hbox{$<$}\kern-.85em
          \lower.35em\hbox{$\sim$}~}\xspace}

\def\pt         {\mbox{$p_{\rm T}$}\xspace}
\def\ptsq         {\mbox{$p_{\rm T}^2$}\xspace}
\def\ptp         {\mbox{$p_{{\rm T}(p)}$}\xspace}
\def\ptpsq         {\mbox{$p_{{\rm T}(p)}^2$}\xspace}

\def\pythia     {\mbox{\textsc{Pythia}}\xspace}

\def\geant      {\mbox{\textsc{Geant4}}\xspace}

\def\tell1  {TELL1\xspace}
\def\ukl1   {UKL1\xspace}

\newcommand{\ie}{\mbox{\itshape i.e.}}

\usepackage{cite} 
\usepackage{mciteplus}

\begin{document}

\renewcommand{\thefootnote}{\fnsymbol{footnote}}
\setcounter{footnote}{1}


\begin{titlepage}
\pagenumbering{roman}

\vspace*{-1.5cm}
\centerline{\large EUROPEAN ORGANIZATION FOR NUCLEAR RESEARCH (CERN)}
\vspace*{1.5cm}
\hspace*{-0.5cm}
\begin{tabular*}{\linewidth}{lc@{\extracolsep{\fill}}r}
\ifthenelse{\boolean{pdflatex}}
{\vspace*{-2.7cm}\mbox{\!\!\!\includegraphics[width=.14\textwidth]{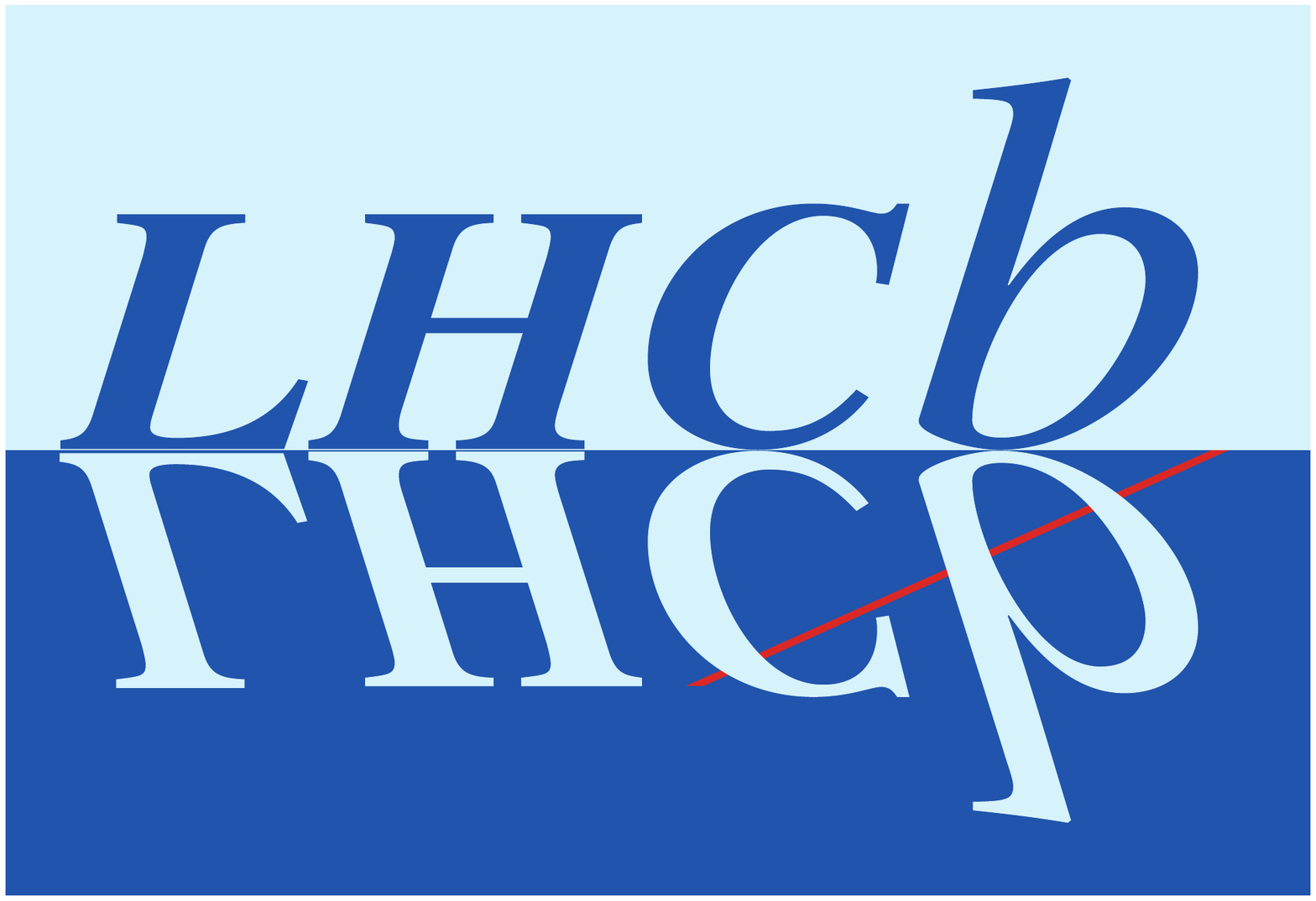}} & &}%
{\vspace*{-1.2cm}\mbox{\!\!\!\includegraphics[width=.12\textwidth]{lhcb-logo.eps}} & &}%
\\
 & & CERN-PH-EP-2014-174 \\  
 & & LHCb-PAPER-2014-027 \\  
 & & 22 July 2014 \\ 
 & & \\
\end{tabular*}

\vspace*{1.5cm}

{\bf\boldmath\huge
\begin{center}
Observation of charmonium pairs produced exclusively 
in  $pp$ collisions
\end{center}
}

\vspace*{0.0cm}

\begin{center}
The LHCb collaboration\footnote{Authors are listed at the end of the paper.}
\end{center}

\vspace*{0.0cm}

\begin{abstract}
 \noindent
A search is performed for the central exclusive production of 
pairs of charmonia produced in proton-proton collisions.
Using data corresponding to an integrated luminosity of
 3\invfb collected at centre-of-mass energies of 7 and 8\tev,
$J/\psi J/\psi$ and $J/\psi\psi(2S)$ pairs are observed, which
have been produced in the absence
of any other activity inside the LHCb acceptance
that is sensitive to charged particles in the pseudorapidity ranges
$(-3.5,-1.5)$ and $(1.5,5.0)$.
Searches are also performed for pairs of P-wave charmonia and
limits are set on their production.
The cross-sections for these  processes, 
where the dimeson system has a rapidity between 2.0 and 4.5,
are measured to be
$$
\begin{array}{rl}
\sigma^{\jpsi\jpsi} &= 58\pm10{(\rm stat)} \pm 6{(\rm syst)} \pb , \\
\sigma^{\jpsi\psitwos} &= 63 ^{+27}_{-18}{(\rm stat)}\pm 10{(\rm syst)} \pb , \\
\sigma^{\psitwos\psitwos} &< 237\pb, \\
\sigma^{\chiczero\chiczero} &< 69\nb, \\
\sigma^{\chicone\chicone} &< 45\pb, \\
\sigma^{\chictwo\chictwo} &< 141\pb, \\
\end{array}
$$
where the upper limits are set at the 90\% confidence level.
The measured $\jpsi\jpsi$ and $\jpsi\psitwos$
cross-sections are consistent with theoretical expectations.

\end{abstract}

\vspace*{1.0cm}

\begin{center}
  Submitted to Journal of Physics G
  \end{center}

\vspace{\fill}

{\footnotesize
\centerline{\copyright~CERN on behalf of the \lhcb collaboration, license \href{http://creativecommons.org/licenses/by/4.0/}{CC-BY-4.0}.}}
\vspace*{2mm}

\end{titlepage}


\newpage
\setcounter{page}{2}
\mbox{~}


\cleardoublepage


\renewcommand{\thefootnote}{\arabic{footnote}}
\setcounter{footnote}{0}



\pagestyle{plain} 
\setcounter{page}{1}
\pagenumbering{arabic}


%

\section{Introduction}

Central exclusive production (CEP), $pp\rightarrow pXp$, in which the protons remain intact  and
the system $X$ is produced with a rapidity gap on either side, requires the exchange
of colourless propagators, either photons or combinations of gluons that ensure a net neutral
colour flow.
CEP provides an attractive laboratory in which to study
quantum chromodynamics (QCD) and the role of the pomeron,
particularly when the mass of the central system is high enough to allow perturbative calculations~\cite{albrow_forshaw}.
Furthermore, it presents an opportunity to search for
exotic states in a low-background experimental environment.

CEP has been studied at hadron colliders from the ISR to the Tevatron. 
At the LHC, measurements of exclusive single \jpsi
photoproduction have been made by the LHCb~\cite{lhcbj} and ALICE~\cite{alice} collaborations.
The CEP of vector meson pairs has been measured in $\omega\omega$~\cite{omegapair}
and $\phi\phi$~\cite{wa102phiphi,phipair} channels by the WA102 and WA76 collaborations.
In this paper, CEP of S-wave, $\jpsi\jpsi,\jpsi\psitwos$, $\psitwos\psitwos$,
and P-wave, $\chiczero\chiczero,\chicone\chicone,\chictwo\chictwo$, charmonium pairs 
are examined for the first time, using a data sample corresponding to 
an integrated luminosity of about 3\invfb, collected by the LHCb experiment.

Investigations of the cross-sections and invariant mass spectra of charmonium pairs
are sensitive to the presence of additional particles in the decay chain such
as glueballs or tetraquarks~\cite{gg_bere}.  
LHCb has measured the inclusive production of \jpsi pairs~\cite{lhcbjj} 
in broad agreement with the QCD predictions,
although the invariant mass distribution of the dimeson system is shifted to higher values in data.
In the inclusive case, this shift could be
an indication of double parton scattering (DPS) effects~\cite{djpsi_stirling}.
In CEP however, DPS through photoproduction
is negligible due to the peripheral nature of the collision.
Thus, a comparison of the mass spectra in inclusive and exclusive production
gives further information for understanding \jpsi pair production.

\begin{figure}[b]
\begin{center}
\includegraphics[width=0.48\linewidth]{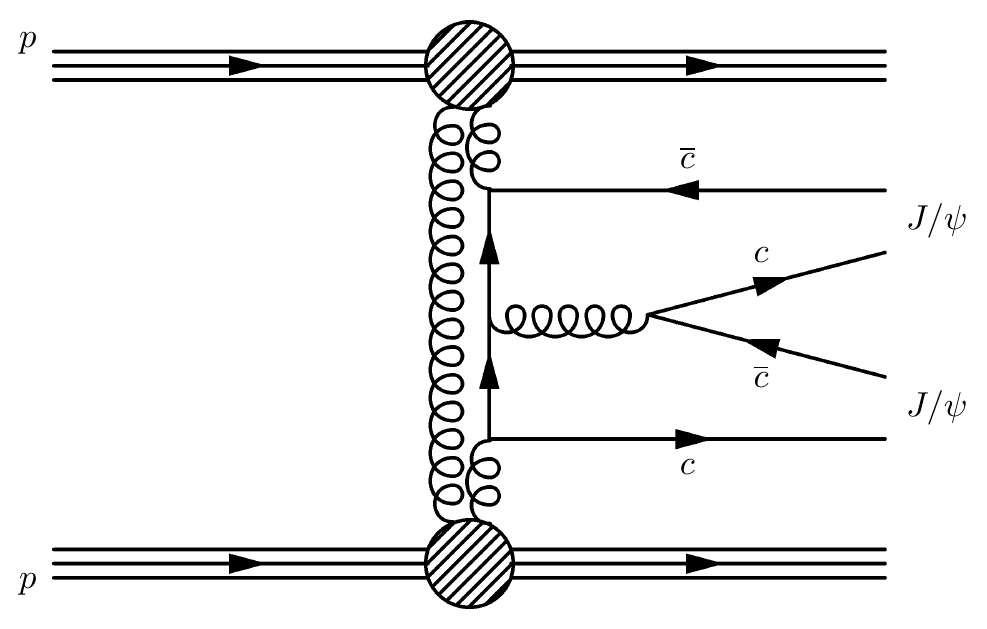}
\includegraphics[width=0.49\linewidth]{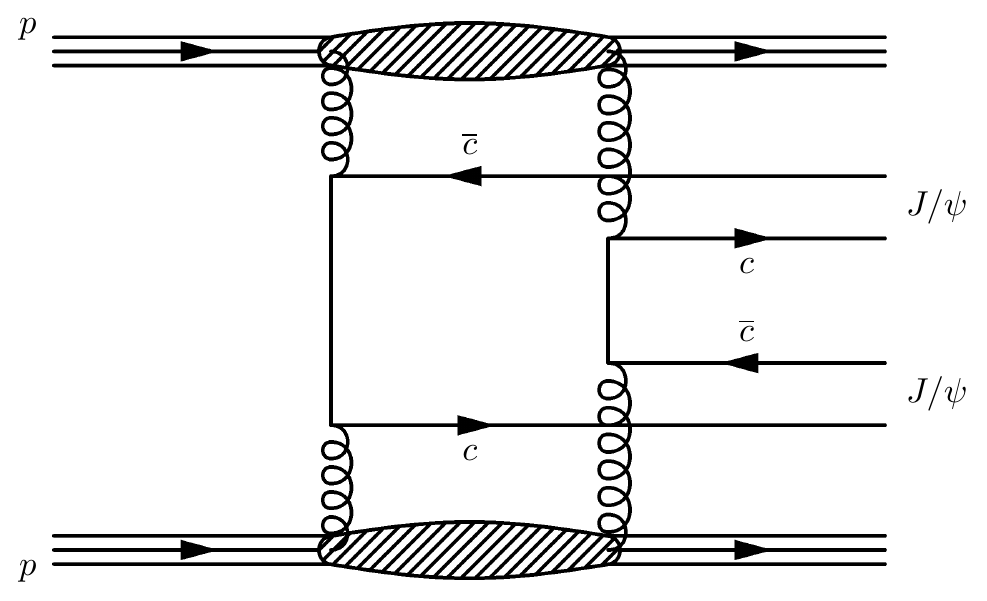}
     \vspace*{-1.0cm}
\end{center}
\caption{\small
Representative Feynman diagrams for pairs of charmonia produced through double pomeron exchange.  In the left, one $t$-channel gluon is much softer than the other while in the right,
they are similar.
}
\label{fig:fd}
\end{figure}

The principal production mechanism for the CEP of two charmonia
is through double
pomeron exchange (DPE) as shown in the left diagram of 
Fig.~\ref{fig:fd}, where one t-channel gluon participates in the hard interaction and the second
(soft) gluon shields the colour charge.
Using the Durham model~\cite{durham_review}, this
can be related to the $gg\rightarrow\jpsi\jpsi$ process 
calculated in Refs.~\cite{gg_qiao,gg_bere}.
Another mechanism~\cite{lucian} that may lead to higher dimeson masses and
an enhanced cross-section is shown in the right diagram of Fig.~\ref{fig:fd}.

Several theory papers consider the production of pairs of charmonia by 
two-photon fusion~\cite{grayson,serbo,qiao,goncalves,antoni,baranov}, 
which is of importance in heavy-ion collisions and at high-energy $e^+e^-$ colliders.
However, in $pp$ interactions DPE dominates.  A recent work~\cite{lucian} gives
predictions for the DPE production of light meson pairs, which are implemented in the
{\sc SuperCHIC} generator~\cite{superCHIC}.  
The formalism can be extended to obtain predictions for charmonium pairs.


\section{Detector and data samples}
\label{sec:data}

The \lhcb detector~\cite{lhcbdet} is a single-arm forward
spectrometer covering the \mbox{pseudorapidity} range $2<\eta <5$ (forward region), 
primarily designed
for the study of particles containing \bquark or \cquark quarks. 
The detector includes a high-precision tracking system consisting of a
silicon-strip vertex detector (VELO)~\cite{velo} surrounding the $pp$ interaction region,
a large-area silicon-strip detector located upstream of a dipole
magnet with a bending power of about $4{\rm\,Tm}$, and three stations
of silicon-strip detectors  and straw drift tubes~\cite{ot} placed
downstream of the magnet.
The tracking system provides a measurement of momentum 
with a relative uncertainty 
that varies from 0.4$\%$ at low momentum
to 0.6$\%$ at 100$\gev$.~\footnote{Natural units are used throughout this paper.}
The minimum distance of a track to a primary vertex, the impact parameter, is measured with a resolution of $(15+29/\pt)\mum$,
where \pt is the component of momentum transverse to the beam, in \gev.
In addition, the VELO has sensitivity to charged particles with momenta above $\sim$100$\mev$ in the \mbox{pseudorapidity} range
 $-3.5<\eta <-1.5$ (backward region), while extending the sensitivity of the forward region to $1.5<\eta<5$.

Different types of charged hadrons are 
distinguished using information from two
ring-imaging Cherenkov detectors~\cite{lhcbrich}. Photon, electron and hadron
candidates are identified by a calorimeter system consisting of
scintillating-pad (SPD) and pre-shower detectors, an electromagnetic
calorimeter and a hadronic calorimeter. The SPD also provides a measure of the charged particle multiplicity in an event.
Muons are identified by a system composed of alternating layers of iron and multiwire
proportional chambers~\cite{lhcbmuon}. 
The trigger~\cite{lhcbtrig} consists of a hardware stage, based
on information from the calorimeter and muon systems, followed by a
software stage, which applies a full event reconstruction.

The data used in this analysis correspond to an integrated luminosity of {$946\pm33$}\invpb
collected in 2011 at a centre-of-mass energy $\sqrt{s}=7\tev$ and {$1985\pm69$}\invpb collected in 2012 at $\sqrt{s}=8\tev$.
The two datasets are combined because the overall yields are low and the cross-sections
are expected to be similar at the two energies.
The $\jpsi$ and $\psitwos$ mesons are identified through their decays to two muons, while the
$\chi_c$ mesons are searched for in the decay channels $\chi_c\rightarrow\jpsi\gamma$. 
The protons are only marginally deflected by the peripheral collision
and remain undetected inside the beam pipe. 
Therefore, the signature for exclusive charmonium pairs is an event containing four muons,
at most two photons,  and no other activity. 
Beam-crossings with multiple proton interactions produce additional activity; 
in the 2011 (2012) data-taking
period the average number of visible interactions per bunch crossing was 1.4 (1.7).
Requiring an exclusive signature
restricts the analysis to beam crossings with a single $pp$ interaction. 

Simulated events are used primarily to determine the detector acceptance. 
No generator has implemented exclusive \jpsi pair production; therefore,
the dimeson system is constructed 
with the mass and transverse momentum distribution
observed in the data, and the rapidity distribution as predicted for DPE processes 
by the Durham model~\cite{durham_review}.
Systematic uncertainties associated with this procedure are discussed in Sec.~\ref{sec:eff}.
The dimeson system is
forced to decay, ignoring spin and polarisation effects,
using the \pythia generator~\cite{pythia}
and
passed through a \geant~\cite{geant} based detector simulation, the trigger emulation and the event reconstruction chain of the $\lhcb$ experiment.

\begin{figure}[b]
\begin{center}
\includegraphics[width=0.49\linewidth]{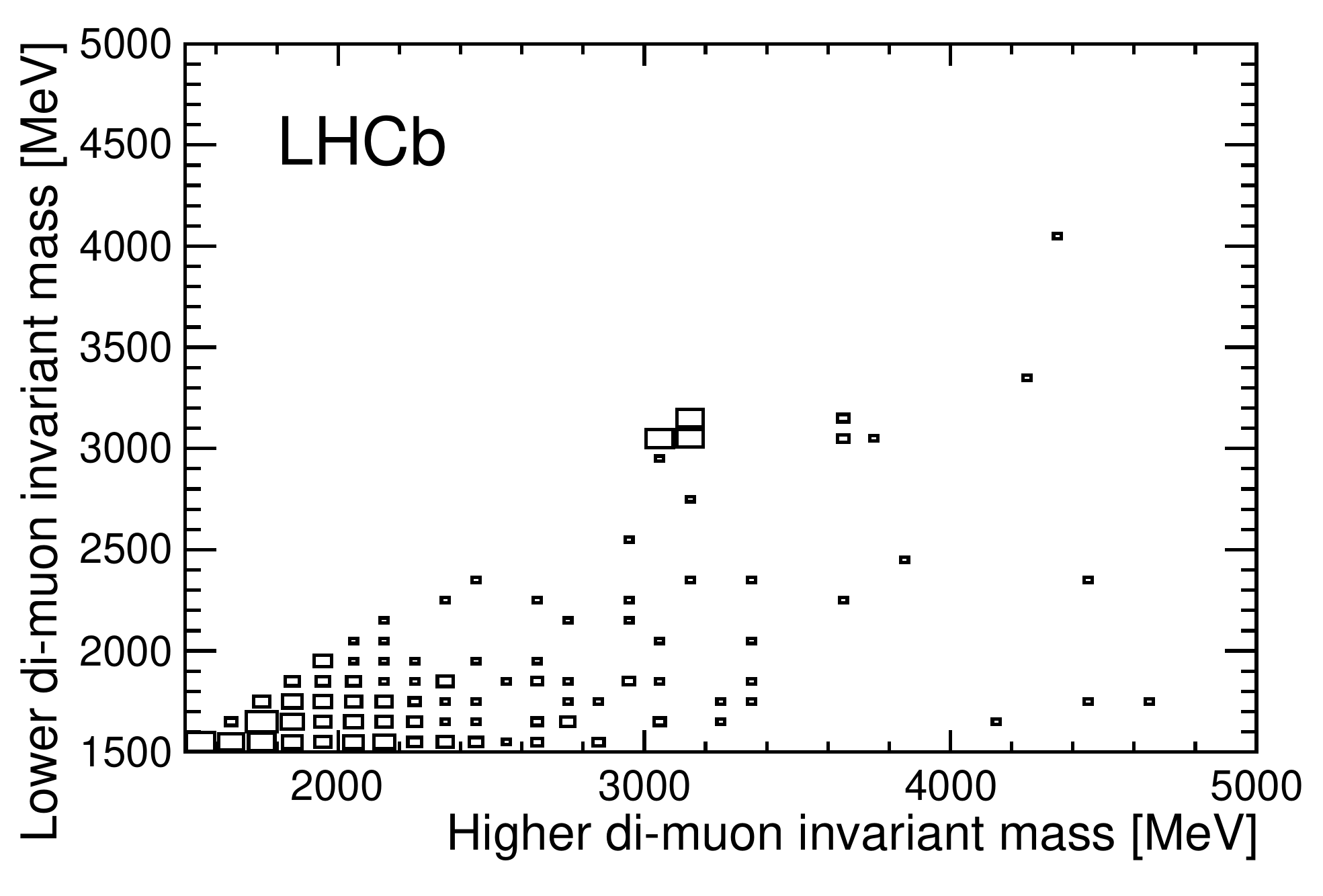}
\includegraphics[width=0.49\linewidth]{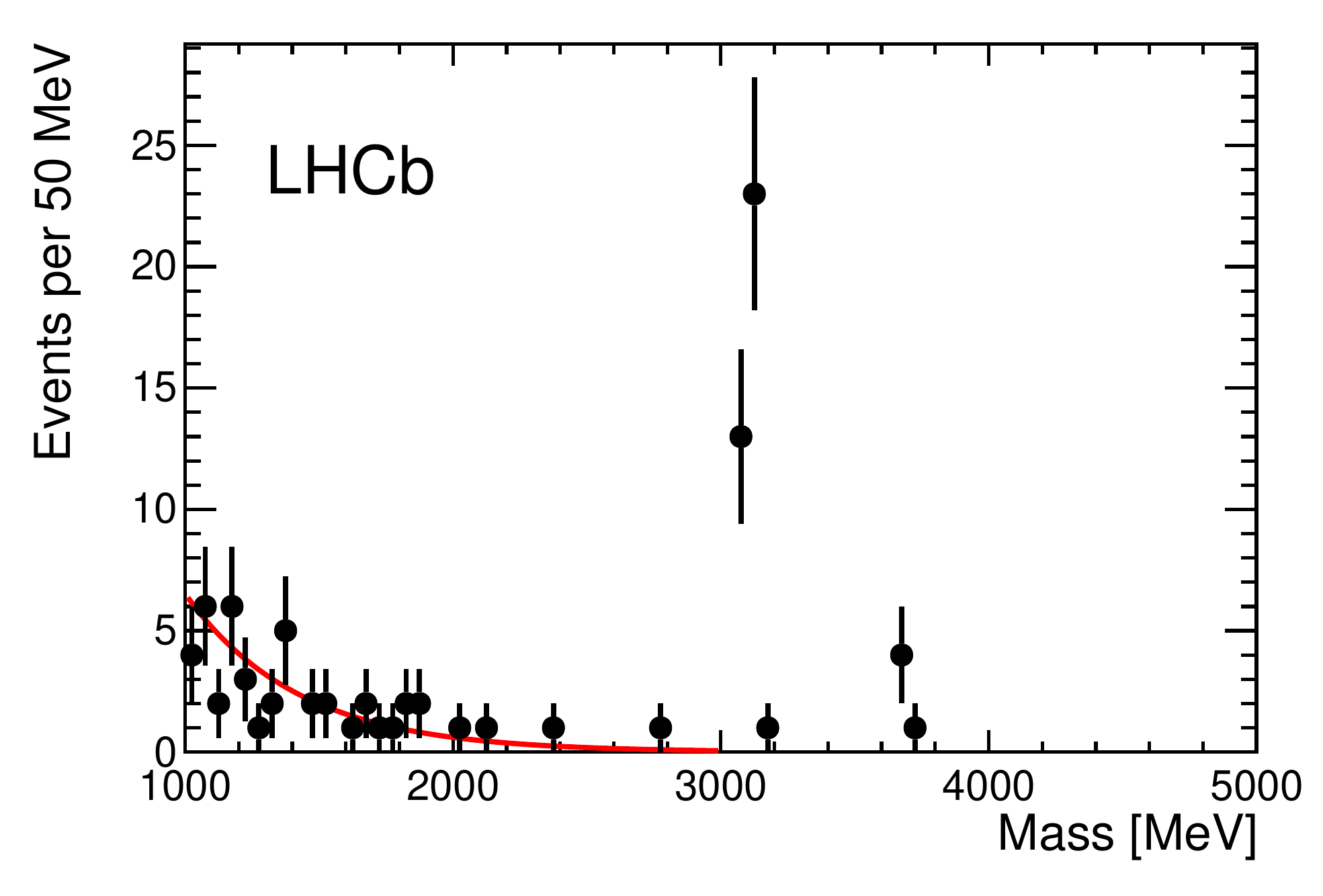}
     \vspace*{-1.0cm}
\end{center}
\caption{\small \small 
Left: Invariant masses of pairs of oppositely charged 
muons in events with exactly four tracks. 
Of the two possible ways of combining the muons per event, 
the one with the higher value for the lower-mass pair is plotted.
Right:  Invariant mass of the second pair of tracks where the first pair
has a mass consistent with the \jpsi or \psitwos meson. 
When both masses are consistent with a charmonium, only the candidate with
the higher mass is displayed.
The curve shows an exponential fit in the region below 2500\mev.
}
\label{fig:mass2}
\end{figure}

\section{Event selection and yields}
\label{sec:sel}

The hardware trigger used in this analysis requires a single muon candidate 
with transverse momentum $\pt$ $>$ 400$\mev$ in coincidence 
with a low SPD multiplicity ($<$ 10 hits).
The software trigger used to select signal events requires two muons with $\pt>400$ \mev.

The analysis is performed in the fiducial region where the dimeson system has a rapidity between 2.0 and 4.5.
The selection of pairs of S-wave charmonia
begins by requiring four reconstructed tracks that incorporate VELO information,
for which the acceptance is about 30\%.  
At least three tracks are required to be identified as muons.
It is required that there are no photons reconstructed in the detector and no other tracks
that have VELO information.

\begin{figure}[t]
\begin{center}
\includegraphics[width=0.49\linewidth]{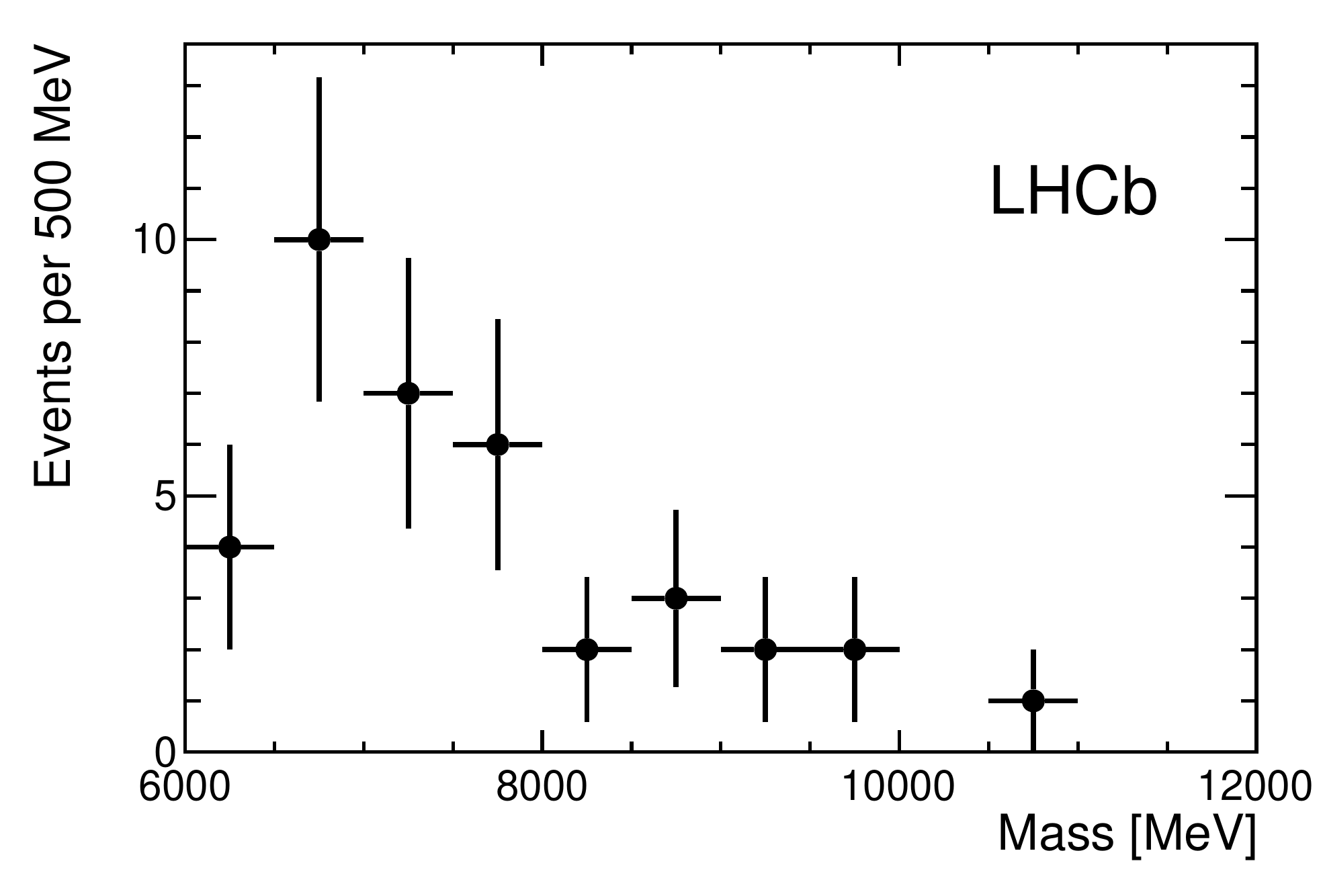}
\includegraphics[width=0.49\linewidth]{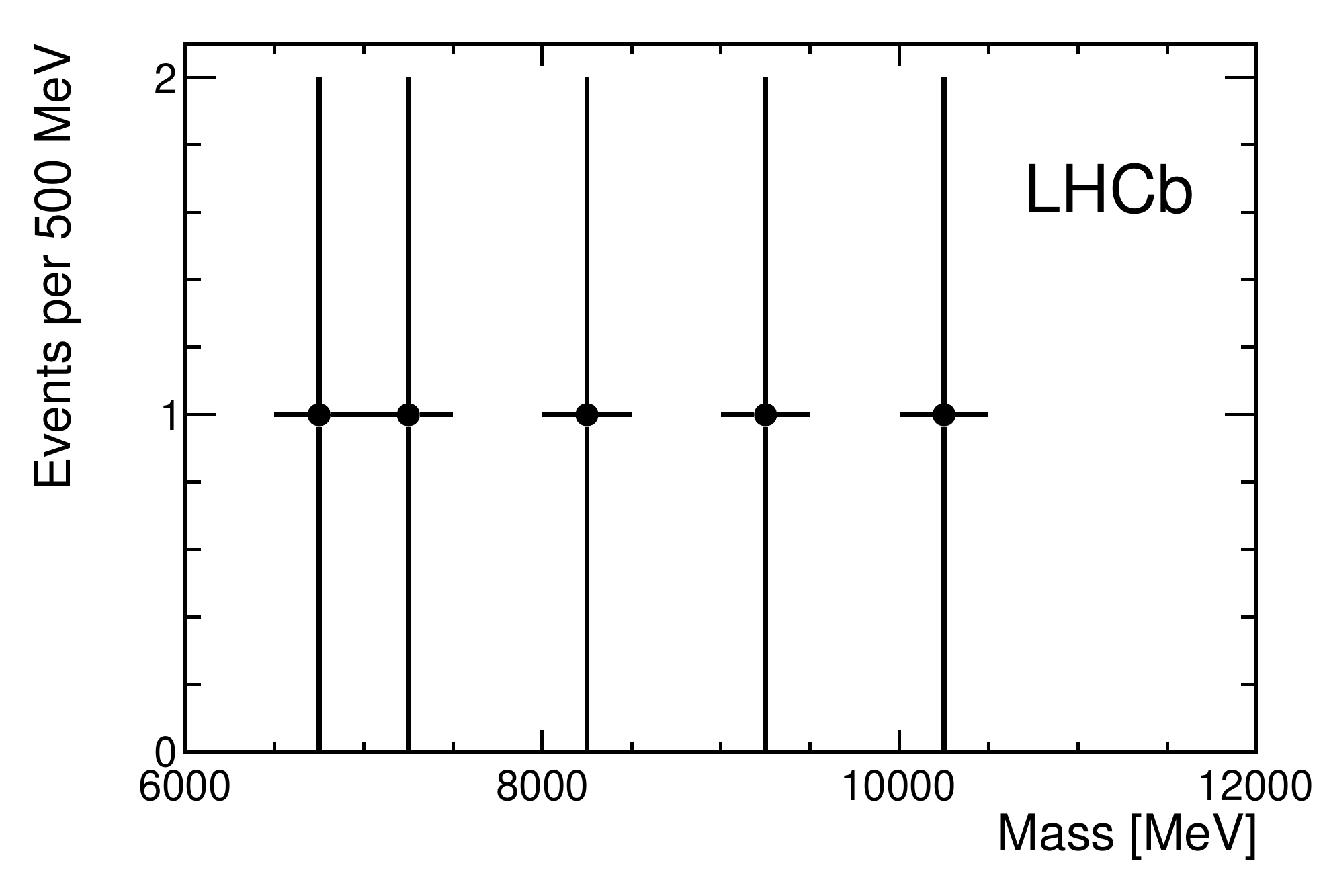}
     \vspace*{-1.0cm}
\end{center}
\caption{\small 
Invariant mass of the four-muon system  in
(left) $\jpsi\jpsi$ and (right) $\jpsi\psitwos$ events.
}
\label{fig:mass4}
\end{figure}

The invariant masses of oppositely charged muon candidates is shown in the left plot of
Fig.~\ref{fig:mass2}.
Accumulations of events are apparent around the \jpsi and \psitwos masses.
Requiring that one of the masses is
within $-200\mev$ and $+65\mev$ of the known \jpsi or \psitwos mass~\cite{pdg}, 
the invariant mass of the other two tracks is shown in the right plot of Fig.~\ref{fig:mass2}.
Clear signals are observed about the \jpsi and \psitwos masses and candidates
within $-200\mev$ and $+65\mev$ of their masses are selected.
There are 37 $\jpsi\jpsi$ candidates, 5 $\jpsi\psitwos$ candidates,
and no $\psitwos\psitwos$ candidates.
Although it is not explicitly required in the selection, all candidates are consistent with 
originating from a single vertex.
The invariant mass distributions of the four-muon system in $\jpsi\jpsi$ and $\jpsi\psitwos$ events
are shown in Fig.~\ref{fig:mass4}.
The shape of the $\jpsi\jpsi$ mass distribution is consistent with that observed in the inclusive
analysis~\cite{lhcbjj}.

\begin{figure}[b]
\begin{center}
\includegraphics[width=0.49\linewidth]{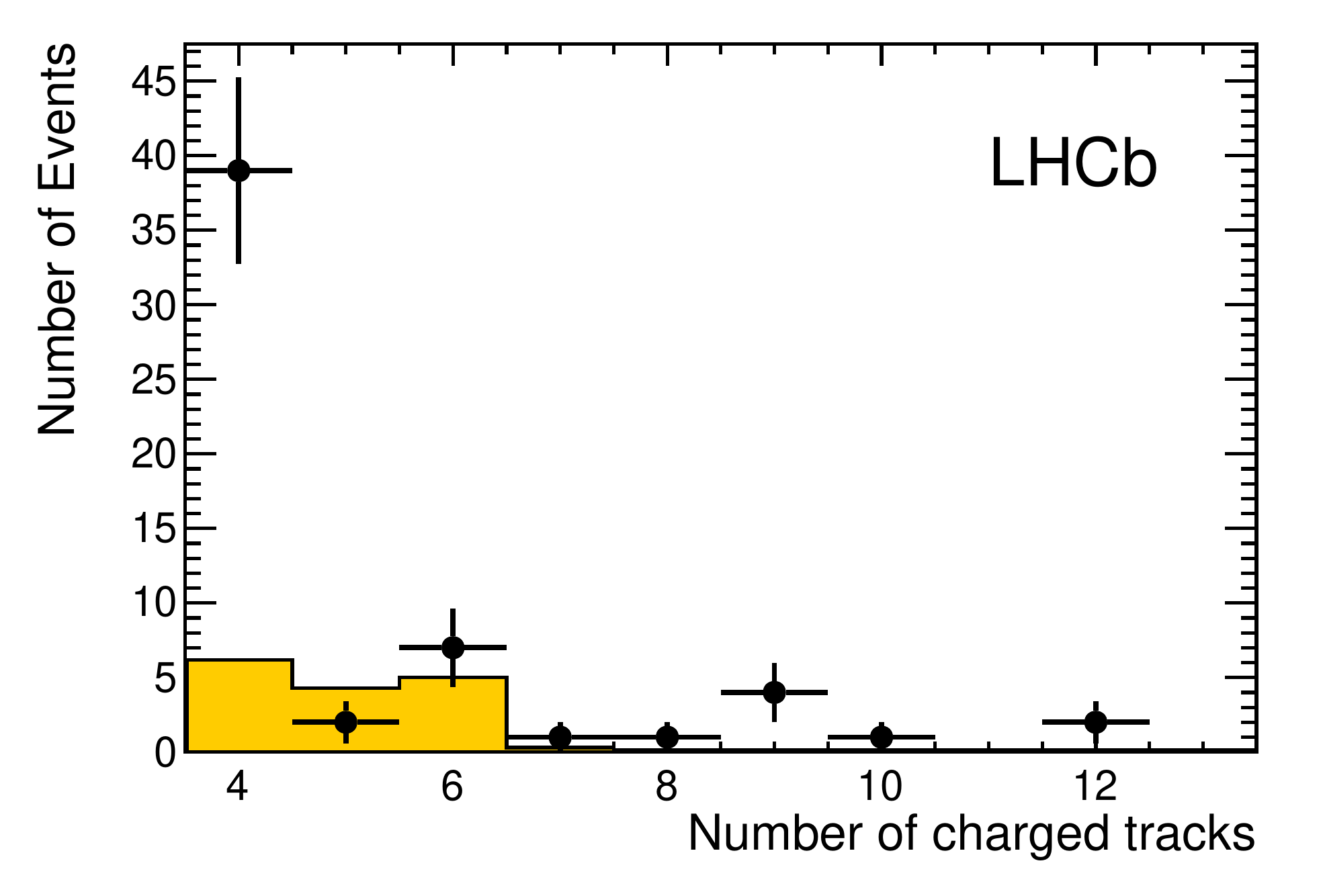}
     \vspace*{-1.0cm}
\end{center}
\caption{\small 
Number of tracks passing the $\jpsi\jpsi$ exclusive selection 
after having removed the requirement that
there be no additional charged tracks or photons.  The shaded histogram is the expected
feed-down from exclusive $\jpsi\psitwos$ events.
}
\label{fig:ex_n}
\end{figure}

The events selected here are produced through a different production mechanism than those selected in the inclusive analysis of \jpsi pairs, as can be appreciated by examining the charged multiplicity distributions.
The inclusive signal has an average multiplicity of 190 reconstructed tracks, with
only 2 (0.2)\% of events having multiplicities below 50 (20).  
In contrast, Fig.~\ref{fig:ex_n} shows the number of tracks, in triggered events
with a low SPD multiplicity, for the selection of
exclusive $\jpsi\jpsi$ events 
when the requirements on no additional activity (either extra tracks or photons)
is removed.
The peak at four tracks is noteworthy.
A small peak of 7 events with six tracks is consistent with the expected
number of
exclusive $\jpsi\psitwos$ events, where $\psitwos\rightarrow\jpsi\pi^+\pi^-$.
This is estimated from the simulation that has been normalised to the 
5 observed $\jpsi\psitwos$ events, where $\psitwos\rightarrow\mu^+\mu^-$.
Only one of these events can be fully
reconstructed and the invariant mass of 
one \jpsi meson and the two tracks, assumed to be pions,
is consistent with that of the \psitwos meson.
The remainder of the distribution is uniform, suggestive of DPE events in which one or both protons dissociated.  There is no indication of a contribution that increases towards higher multiplicities, as
would be expected if there was a substantial contribution coming from non-exclusive events.

The selection of pairs of P-wave charmonia proceeds as for the S-wave, but the
restriction on the number of photons is lifted.  These criteria are only satisfied by two events.
One event has a single photon and the invariant mass of a reconstructed
\jpsi and this photon is consistent with the $\chiczero$ mass;
consequently, this event is a candidate for $\chiczero\chiczero$ production.  
The other event has two photons that, when combined,
have the mass of a $\pi^0$ meson, and is thus not a candidate for $\chi_c\chi_c$ production.
Both events are consistent with 
partially reconstructed $\jpsi\psitwos$ events where $\psitwos\rightarrow\jpsi\pi^0\pi^0$.
Normalising to the five candidate events for $\jpsi\psitwos$, 
the simulation estimates that
$2.8\pm2.0\ (0.5\pm0.5)$ 
$\jpsi\psitwos$ events would be reconstructed 
as $\jpsi\jpsi$ candidates with one (two) additional photon(s).
There are no candidates for $\chicone\chicone$ or $\chictwo\chictwo$ production.

\section{Backgrounds}
\label{sec:bkg}

Three background components are considered: non-resonant background;  
feed-down from the exclusive production of other mesons; 
and inelastic production of mesons where one or both protons dissociate.

The non-resonant background is only considered for the
S-wave analysis and is calculated by fitting an exponential to the non-signal
contribution in Fig.~\ref{fig:mass2} and extrapolating under the signal.  
It is estimated that 
there are $0.3\pm0.1$ and $0.07\pm0.02$ background events  in the \jpsi and \psitwos 
signal ranges, respectively.

A feed-down background is considered for the $\jpsi\jpsi$ and the P-wave analyses.
Given the presence of five $\jpsi\psitwos$ signal events, 
it is expected that 
$\psitwos\rightarrow\jpsi X$ decays will occasionally be reconstructed as $\chi_c$ mesons or
$\jpsi$ mesons alone, 
due to the rest of the decay products being outside the acceptance or below threshold.
Normalising to the five candidate events for $\jpsi\psitwos$, 
the simulation estimates that
$2.9\pm2.0$ $\jpsi\psitwos$ events would be reconstructed 
as $\jpsi\jpsi$ candidates with no additional photons, while
$0.8\pm0.8$,
$0.2\pm0.2$ and
$0.1\pm0.1$
would be reconstructed as
$\chiczero,\chicone$ and $\chictwo$ mesons, respectively.

Feed-down from pairs of P-wave charmonia to give $\jpsi\jpsi$ candidates
is also possible.
The simulation estimates that in over 
80\% (70\%) of $\chicone\chicone$ or $\chictwo\chictwo$ 
($\chiczero\chiczero$) decays producing two $\jpsi$ mesons, one or more
additional photons would be detected.
There is only one candidate for $\chiczero\chiczero$ but this is also consistent with
feed-down from $\jpsi\psitwos$ events.  Consequently, there is  
no evidence for a significant $\chi_c$ feed-down to the $\jpsi\jpsi$ selection
and this contribution is assumed to be negligible.

\begin{figure}[b]
\begin{center}
\includegraphics[width=0.49\linewidth]{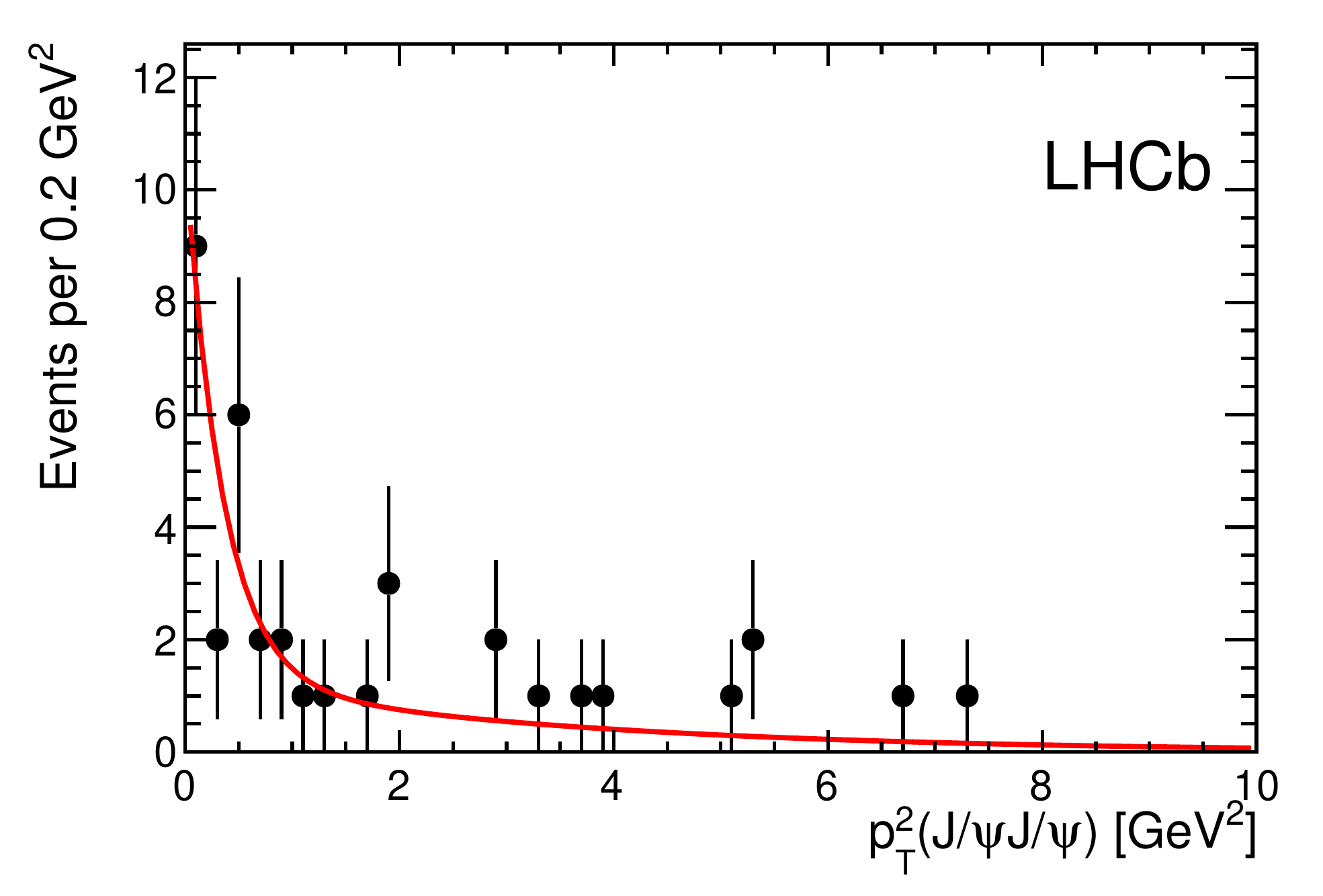}
\includegraphics[width=0.49\linewidth]{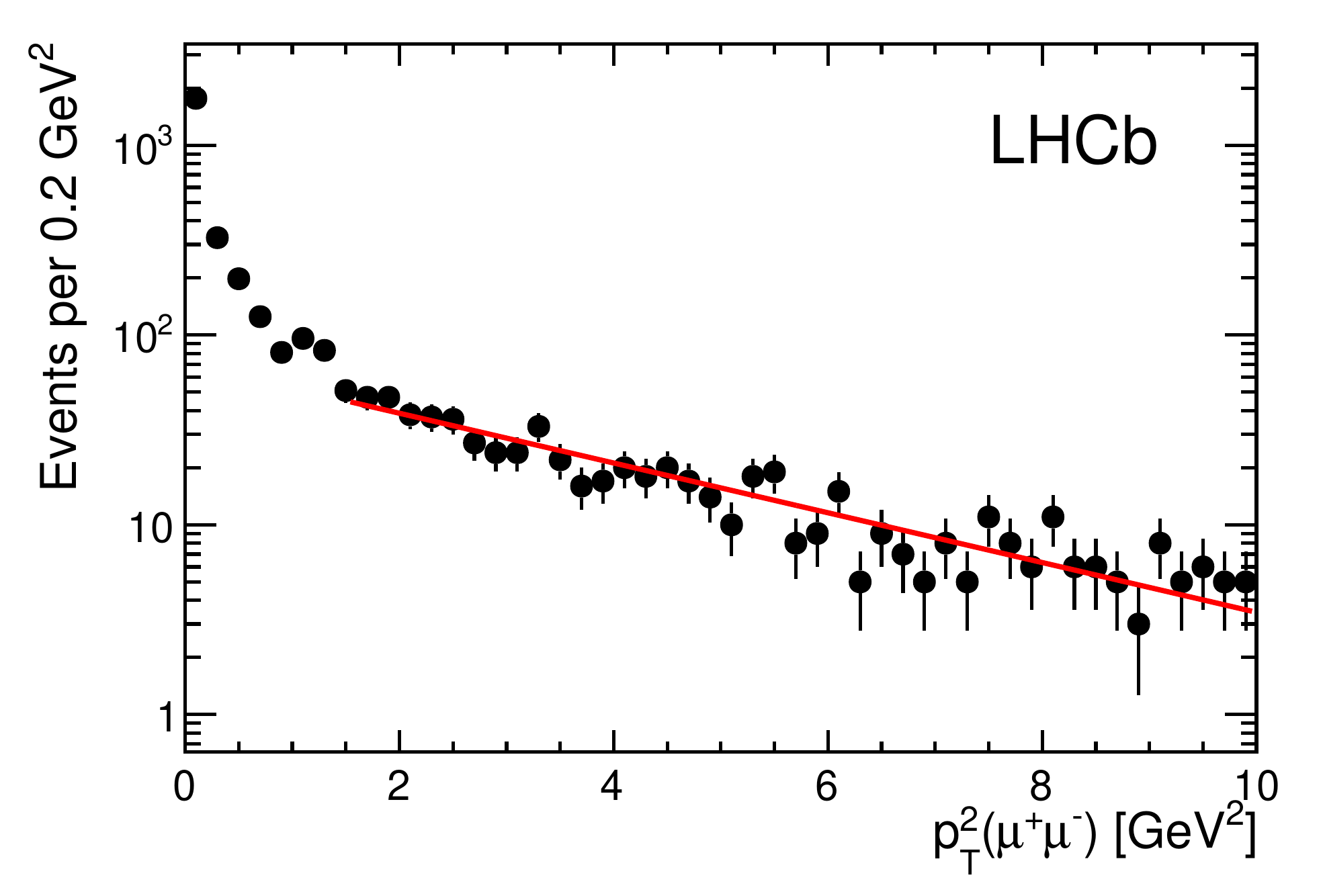}
     \vspace*{-1.0cm}
\end{center}
\caption{\small 
Transverse momentum squared distribution of candidates for exclusively produced  
(left) $\jpsi\jpsi$ and
(right) dimuons whose invariant mass is between 6 and 9 \gev.
The curves are fits to the data as described in the text.
}
\label{fig:pt2}
\end{figure}

The separation of the samples into those events that are truly exclusive (elastic) 
and those where one
or both protons dissociate (inelastic) 
is fraught with difficulty.
Therefore, the cross-sections
are quoted for the full samples
that are observed to be exclusively produced inside the LHCb acceptance, 
\ie\  no other tracks or electromagnetic deposits are found in the detector.  
Nonetheless, to compare with theoretical predictions that
are usually quoted for the elastic process without proton break-up, an attempt is made to
quantify the elastic fraction in the  $\jpsi\jpsi$ sample, 
using the distribution of squared transverse momentum 
and describing the elastic and proton-dissociation components
by different exponential functions.
This functional form is suggested by Regge theory
that assumes the differential cross-section
$d\sigma/ dt\propto\exp(bt)$
for a wide class of diffractive events, where $b$ is a constant for a given process,
$t\approx -\ptpsq$ is 
the four-momentum transfer squared at one of the proton-pomeron vertices,
and $\ptp$ is the transverse momentum of the outgoing proton labelled $(p)$.

In the CEP single \jpsi analysis performed by the LHCb collaboration~\cite{lhcbj}, 
the transverse momentum of the central system,
$\pt\approx\ptp$, the transverse momentum of the outgoing proton from which the pomeron radiated.
A fit to the \ptsq distribution showed that the elastic contribution could be described by
$d\sigma/ d\ptsq\sim\exp(-(6\gev^{-2})\ptsq)$ and it was estimated
that about 60\% of events with $\ptsq<1\gev^2$ 
(corresponding to 40\% of events without a requirement on \pt)
were elastic.

In the CEP of pairs of  \jpsi mesons, the situation should be similar, although
$d\sigma/ d\ptsq$ will fall off more gradually as there are two proton-pomeron vertices to
consider.
In addition, the dissociative background might be larger as the mass of the central system
is higher and the production process is through
two pomerons, rather than a photon and a pomeron.  
The transverse momentum of the central system,
 $\ptsq=\ptsq_{(p1)}+\ptsq_{(p2)}+2\vec{p}_{{\rm T}(p1)}\cdot\vec{p}_{{\rm T}(p2)}$.
 Taking a dependence of $\exp{(-(6\gev^{-2})\ptpsq)}$ at each of the proton-pomeron vertices 
 and ignoring possible rescattering effects leads
 to an expectation of 
$d\sigma/ d\ptsq\sim\exp(-(3\gev^{-2})\ptsq)$.
The $\ptsq$ distribution for the $\jpsi\jpsi$ candidates
is shown in the left plot of Fig.~\ref{fig:pt2}
and has a shape similar to that seen in the exclusive \jpsi analyses:
a peaking of signal events below
$1 \gev^2$, and a tail to higher values, characteristic of inelastic production.
A maximum likelihood fit is performed to the sum of two exponentials,
\begin{equation}
f_{\rm el}b_s\exp{(-b_s \ptsq)}+{(1-f_{\rm el})b_b}\exp{(-b_b\ptsq)},
\label{eq:pur}
\end{equation}
where $b_s,b_b$ are the slopes for the signal and background and $f_{\rm el}$ is the 
fraction of elastic events.

Due to the small sample size, the value of $b_b$ is constrained using the
distribution for exclusive dimuon candidates whose invariant mass lies in the range 6 to 9\gev.
These are selected as in the single \jpsi analysis~\cite{lhcbj} but with a different invariant mass
requirement.  
The $\ptsq$ distribution, shown in the right plot of Fig.~\ref{fig:pt2}, 
has a prominent peak in the first bin corresponding to 
the electromagnetic two-photon exchange process, $pp\rightarrow p\mu^+\mu^- p$.
The tail to larger values is characteristic of events
with proton dissociation.
The region $1.5<\ptsq<10\gev^2$ is fit with a single exponential, resulting in a slope
of $b_b=0.29\pm0.02\gev^{-2}$. 

Fixing $b_b$ at this value of $0.29\gev^{-2}$, the fit to the $\ptsq$ distribution for
the $\jpsi\jpsi$ candidates
returns values of $b_s=2.9\pm1.3\gev^{-2}$ and $f_{\rm el}=0.42\pm0.13$.
An alternative fit is made with all parameters free, returning consistent results, albeit with
larger uncertainties: 
$b_s=3.1\pm1.7\gev^{-2}$,
$b_b=0.34\pm0.14\gev^{-2}$,
$f_{\rm el}=0.38\pm0.17$.
It is also worth noting that the $\ptsq$ spectrum of these selected
events is different to that of inclusively selected \jpsi pairs, which can be fit with a single
exponential with a slope of $0.051\pm0.001\gev^{-2}$.


\section{Efficiency and acceptance}
\label{sec:eff}

For dimesons in the rapidity range $2.0<y<4.5$, 
the acceptance factor, $A$, defines the fraction of events having four 
reconstructed tracks in the LHCb detector.
This is  found using simulated events that have been generated with a smoothed
form of the distribution
given in the left plot of Fig.~\ref{fig:mass4}, 
a $\ptsq$ as given in the left plot of Fig.~\ref{fig:pt2},
and a rapidity distribution according to the Durham model, which
in the region $2.0<y<4.5$, can be described by the functional form $(1-0.18y)$.

To estimate a systematic uncertainty on the acceptance, the simulated events are reweighted
with different assumptions on the mass, transverse momentum and rapidity of the dimeson
system.
The mass is described instead by the theoretical shape of  Ref.~\cite{gg_bere}, which changes
the acceptance by less than $1$\%.
The transverse momentum is described with the elastic shape found in data for
the exclusive single \jpsi analysis~\cite{lhcbj}, changing the acceptance by less than $1$\%.
The largest effect is due to the assumption on the underlying rapidity distribution.
An alternative model
is to consider the pomeron as an isoscalar photon~\cite{Donn87} and construct the pomeron flux from the
Weizs\"acker-Williams approximation for describing photon radiation.
This leads to a rapidity
distribution that is
approximately flat for $2.0<y<4.5$ and 
an acceptance that changes by 6\%.

The tracking efficiency also contributes to $A$.
A systematic uncertainty of 1\% per track has been determined~\cite{lhcb_track} 
for tracks with pseudorapidities between 2.0 and 4.5.  Uncertainties in the description
of edge effects of the tracking detectors in the simulation are assessed by
comparing the pseudorapidity distributions of tracks in exclusive \jpsi events~\cite{lhcbj} 
in simulation and data.
Differences are propagated to give an uncertainty on the determination of $A$ for
charmonium pairs of 7\%.
Combining all these effects in quadrature 
leads to estimates of $0.35\pm0.03$ for the acceptance of $\jpsi\jpsi$ events,
$0.36\pm0.03$ for the acceptance of $\jpsi\psitwos$ and $\psitwos\psitwos$ events,
and $0.29\pm0.03$ for the acceptance of any of the $\chi_c\chi_c$ pairs.

The efficiency, $\epsilon$, for triggering and reconstructing signal events is the product of
three quantities: $\epsilon_{\rm trigger}$,
$\epsilon_{\rm muid}$
and
$\epsilon_{\rm sel}$.
The trigger only requires two of the four muons and consequently has a high
efficiency of  $\epsilon_{\rm trigger}=0.90\pm0.03$.
This has been calculated from the single muon trigger efficiencies calculated in Ref.~\cite{lhcbj}
together with the efficiency for the SPD multiplicity to be less than ten, 
which has been assessed using a rate-limited trigger that does not have requirements
on the SPD multiplicity. 
The efficiency to identify three or more of the final-state decay products as muons, $\epsilon_{\rm muid}$,
is high
and the uncertainty is determined by propagating the difference in
single muon efficiencies found in simulation and in data to give 
$\epsilon_{\rm muid}=0.95\pm0.03$. 
For the S-wave analysis, the selection has an efficiency $\epsilon_{\rm sel}=0.93\pm0.02$, which
includes contributions from the requirements that
no photons be identified in the event and that the reconstructed masses be within
$-200\mev$ and $+65\mev$ of the \jpsi or \psitwos mass.
The former is found from simulation, calibrated using a sample of $\jpsi\gamma$ candidates in data, 
while the latter is determined from a fit to the peak in the exclusive \jpsi analysis~\cite{lhcbj}.
For the P-wave analysis, the requirement of detecting one or more photons lowers the selection efficiency and values of $\epsilon_{\rm sel}=0.68\pm0.07,0.77\pm0.04,0.81\pm0.04$ are obtained for
$\chiczero\chiczero$, $\chicone\chicone$, $\chictwo\chictwo$, respectively, where the
uncertainty takes into account the modelling of the energy response of the calorimeter.


\section{Results and discussion}
The cross-section, $\sigma^{M_1M_2}$,
for the production of meson pairs, $M_1$ and $M_2$,
is given by 
\begin{equation}
\sigma^{M_1M_2}=
{N_{M_1M_2}-N_{\rm bkg}\over 
(f_{\rm single} L)
\ A\ \epsilon \
{\cal B}({M_1\rightarrow\mu\mu(\gamma)})\ 
{\cal B}({M_2\rightarrow\mu\mu(\gamma)})}
\label{eq:cs}
\end{equation}
where $N_{M_1M_2}$ is the number of candidate meson pairs selected, 
$N_{\rm bkg}$ is the estimated number of background
events, 
$L$ is the integrated luminosity, 
$f_{\rm single}$ is the fraction of beam crossings with a single interaction,
and ${\cal B}({M_i\rightarrow\mu\mu(\gamma)})$ is the branching fraction 
for the meson to decay to two muons in the case of S-wave states,
and two muons and a photon for P-wave states.

\begin{table}
\begin{small}
\begin{center}
\caption{\small Summary of numbers entering the cross-section calculation.}
\begin{tabular}{|ccccccc|}
\hline
 & $\jpsi\jpsi$ & $\jpsi\psitwos$ & $\psitwos\psitwos$ & $\chiczero\chiczero$ & $\chicone\chicone$ & $\chictwo\chictwo$ \\
\hline
$N_{M_1M_2}$ &  37 & 5 & 0 & 1 & 0 & 0\\
$N_{\rm bkg}$   & $3.2\pm2.0$ & $0.07\pm0.02$ & $<0.01$ & $0.8\pm 0.8$ & $0.2\pm0.2$ & $0.1\pm 0.1$ \\
$A$  
& $0.35\pm0.03$
& $0.36\pm0.03$
& $0.36\pm0.03$
& $0.29\pm0.03$
& $0.29\pm0.03$
& $0.29\pm0.03$\\
$\epsilon$ & $0.80\pm0.04$ & $0.80\pm0.04$ & $0.80\pm0.04$ & $0.58\pm0.06$ & $0.66\pm0.05$ & $0.69\pm0.05$\\
\hline
$f_{\rm single}L$[\invpb] & \multicolumn{6}{c|}{ $596\pm 21$} \\
${\cal B}({\jpsi\rightarrow\mu\mu})$ & \multicolumn{6}{c|}{$0.0593\pm0.0006$ }\\
${\cal B}({\psitwos\rightarrow\mu\mu})$  & \multicolumn{6}{c|}{ $0.0077\pm0.0008$} \\
${\cal B}({\chiczero\rightarrow\jpsi\gamma})$  & \multicolumn{6}{c|}{ $0.0117\pm0.0008$} \\
${\cal B}({\chicone\rightarrow\jpsi\gamma})$  & \multicolumn{6}{c|}{ $0.344\pm0.015$} \\
${\cal B}({\chictwo\rightarrow\jpsi\gamma})$  & \multicolumn{6}{c|}{ $0.195\pm0.008$} \\
\hline
\end{tabular}
\end{center}
\label{tab:cs}
\end{small}
\end{table}

\begin{table}
\begin{small}
\caption{\small Percentage uncertainties on the quantities entering the denominator of Eq.\,(\ref{eq:cs}).}
\begin{center}
\begin{tabular}{|ccccccc|}
\hline
Quantity & \multicolumn{6}{c|}{Relative size of systematic uncertainty [\%]} \\
 & $\jpsi\jpsi$ & $\jpsi\psitwos$ & $\psitwos\psitwos$ & $\chiczero\chiczero$ & $\chicone\chicone$ & $\chictwo\chictwo$ \\
 \hline
$A$ & 9 & 9 & 9 & 9 & 9 & 9 \\
$\epsilon$ & 5 & 5& 5 & 10 & 7 & 7  \\
$f_{\rm single}L$ [\invpb] & 3.5 & 3.5 & 3.5& 3.5 & 3.5 & 3.5\\
${\cal B}(M_1){\cal B}(M_2)$
 & 2& 10 & 21& 14 & 8 & 9\\
\hline
Total & 11 & 15 & 24 & 20 & 14 & 15 \\
\hline
\end{tabular}
\end{center}
\end{small}
\label{tab:sys}
\end{table}

The luminosity has been determined with an uncertainty of 3.5\%~\cite{lumi}. 
The factor, $f_{\rm single}$, 
accounts for the fact that the selection requirements reject signal events that are accompanied by a visible proton-proton interaction in the same beam crossing, and is calculated as described in Ref.~\cite{lhcbj}.
The cross-section measurements for 2011 and 2012 data are consistent and are combined to
produce results at an average
centre-of-mass energy of 7.6\tev.
All the numbers entering the cross-section calculation are given in Table~\ref{tab:cs}, while
the systematic uncertainties of the quantities in the denominator of Eq.\,(\ref{eq:cs})
are summarised in Table~\ref{tab:sys}.
Where zero or one candidate is observed, 90\% confidence levels (CL) are calculated 
by performing pseudo-experiments
in which the quantities in Eq.\,(\ref{eq:cs}) are varied according to their uncertainties and
pseudo-candidates are generated according to a Poisson distribution.
The upper bound at 90\% CL is defined as 
the smallest cross-section value that in 90\% of pseudo-experiments leads to
more candidate events than observed in data.
The cross-sections, at an average energy of 7.6\tev, 
for the dimeson system to be in the rapidity range $2.0<y<4.5$ with
no other charged or neutral energy inside the LHCb acceptance are measured to be
\begin{equation*}
\label{eq:result}
\begin{array}{rl}
\sigma^{\jpsi\jpsi} &= 58\pm10{(\rm stat)} \pm 6{(\rm syst)} \pb , \\
\sigma^{\jpsi\psitwos} &= 63 ^{+27}_{-18}{(\rm stat)}\pm 10{(\rm syst)} \pb , \\
\sigma^{\psitwos\psitwos} &< 237\pb, \\
\sigma^{\chiczero\chiczero} &< 69\nb, \\
\sigma^{\chicone\chicone} &< 45\pb, \\
\sigma^{\chictwo\chictwo} &< 141\pb, \\
\end{array}
\end{equation*}
where the upper limits are at 90\% CL.
To compare with theory, the elastic fraction is taken to be
$0.42\pm0.13$, as determined in Sec.~\ref{sec:bkg}, to give an estimated 
cross-section for central exclusive production of $\jpsi\jpsi$ of $24\pm9\pb$, where 
all the uncertainties are combined in quadrature.  
Using the formalism of Ref.\cite{lucian}, a preliminary prediction~\cite{private} of 8\pb at $\sqrt{s}=8\tev$
has been obtained.
There is a large uncertainty of a factor two to three on this value
due to the gluon parton density function  that enters with the fourth power,
the choice of the gap survival factor~\cite{gap},
and the value of the \jpsi wave-function at the origin~\cite{gg_qiao,gg_bere,bodwin}. 
Theory and experiment are observed to be in reasonable agreement,
given the large uncertainties that currently exist on both.

The relative sizes of the cross-sections for exclusive
$\jpsi\psitwos$ and $\jpsi\jpsi$ production, assuming a similar elastic fraction, is 
\begin{equation*}
{
\sigma(\jpsi\psitwos)
\over
\sigma(\jpsi\jpsi)
}
=1.1^{+0.5}_{-0.4},
\end{equation*}
where the total uncertainty is quoted and
most systematics, bar that on the branching fractions, cancel in the ratio.
This is in agreement 
with a theoretical estimate for this ratio of about 0.5~\cite{gg_bere}. 
The equivalent quantity measured in
exclusive single charmonium production~\cite{lhcbj} is
\begin{equation*}
{
\sigma(\psitwos)
\over
\sigma(\jpsi)
}
=0.17\pm0.02.
\end{equation*}
No strong conclusion can be drawn on
the higher relative fraction of \psitwos to \jpsi
in double charmonium production compared to that in single charmonium production,
due to the large uncertainty.

The ratios of double to single  \jpsi production can be compared in the inclusive and exclusive
modes.
The inclusive ratio was measured by LHCb~\cite{lhcbjj} to be
\begin{equation*} 
\sigma^{\jpsi\jpsi}/\sigma^{\jpsi}|_{\rm inclusive}=
(5.1 \pm 1.0 \pm 0.6 ^{+1.2}_{-1.0})\times 10^{-4}.
\end{equation*}
The exclusive single \jpsi cross-section 
is calculated from the differential cross-section and acceptance values
given in Tables 2 and 3 of Ref.~\cite{lhcbj} and the branching fraction to dimuons to get
$\sigma^\jpsi=11.2\pm0.8\nb $.
A combination with 
the estimated exclusive elastic cross-section above 
gives
\begin{equation*}
\sigma^{\jpsi\jpsi}/\sigma^{\jpsi}|_{\rm exclusive}=
(2.1\pm 0.8) \times 10 ^{-3},
\end{equation*} 
the central value for which is four times higher than in the inclusive case, though again,
due  to the large uncertainty, both are consistent.


\section{Conclusions}

A clear signal for the production of pairs of S-wave charmonia
in the absence of other activity in the LHCb acceptance is obtained.
This is the first observation of the central exclusive production of pairs of charmonia.
The small sample size affects the precision with which the elastic component can be extracted.
A fit to the $\ptsq$ of the system of two \jpsi mesons estimates that
$(42\pm13)\%$ of the events that are observed to be exclusive in the LHCb detector
is elastically produced while the remainder is attributed to events in which
one or both protons dissociate.
The measurement are in agreement with preliminary theoretical predictions.
No signal is observed for the production of pairs of P-wave charmonia and upper limits 
on the cross-sections are set.

\section*{Acknowledgements}
  
\noindent 
We thank Lucian Harland-Lang and Valery Khoze for many helpful discussions and for providing
theoretical predictions.
We express our gratitude to our colleagues in the CERN
accelerator departments for the excellent performance of the LHC. We
thank the technical and administrative staff at the LHCb
institutes. We acknowledge support from CERN and from the national
agencies: CAPES, CNPq, FAPERJ and FINEP (Brazil); NSFC (China);
CNRS/IN2P3 (France); BMBF, DFG, HGF and MPG (Germany); SFI (Ireland); INFN (Italy);
FOM and NWO (The Netherlands); MNiSW and NCN (Poland); MEN/IFA (Romania);
MinES and FANO (Russia); MinECo (Spain); SNSF and SER (Switzerland);
NASU (Ukraine); STFC (United Kingdom); NSF (USA).
The Tier1 computing centres are supported by IN2P3 (France), KIT and BMBF
(Germany), INFN (Italy), NWO and SURF (The Netherlands), PIC (Spain), GridPP
(United Kingdom).
We are indebted to the communities behind the multiple open
source software packages on which we depend. We are also thankful for the
computing resources and the access to software R\&D tools provided by Yandex LLC (Russia).
Individual groups or members have received support from
EPLANET, Marie Sk\l{}odowska-Curie Actions and ERC (European Union),
Conseil g\'{e}n\'{e}ral de Haute-Savoie, Labex ENIGMASS and OCEVU,
R\'{e}gion Auvergne (France), RFBR (Russia), XuntaGal and GENCAT (Spain), Royal Society and Royal
Commission for the Exhibition of 1851 (United Kingdom).






\addcontentsline{toc}{section}{References}
\bibliographystyle{LHCb}
\bibliography{ref_pap}

\newpage
\centerline{\large\bf LHCb collaboration}
\begin{flushleft}
\small
R.~Aaij$^{41}$, 
B.~Adeva$^{37}$, 
M.~Adinolfi$^{46}$, 
A.~Affolder$^{52}$, 
Z.~Ajaltouni$^{5}$, 
S.~Akar$^{6}$, 
J.~Albrecht$^{9}$, 
F.~Alessio$^{38}$, 
M.~Alexander$^{51}$, 
S.~Ali$^{41}$
% ...

{\footnotesize \it
% ...
$ ^{5}$Clermont Universit\'{e}, Universit\'{e} Blaise Pascal, CNRS/IN2P3, LPC, Clermont-Ferrand, France\\
$ ^{6}$CPPM, Aix-Marseille Universit\'{e}, CNRS/IN2P3, Marseille, France\\
% ...
$ ^{9}$Fakult\"{a}t Physik, Technische Universit\"{a}t Dortmund, Dortmund, Germany\\
% ...
$ ^{37}$Universidad de Santiago de Compostela, Santiago de Compostela, Spain\\
$ ^{38}$European Organization for Nuclear Research (CERN), Geneva, Switzerland\\
% ...
$ ^{41}$Nikhef National Institute for Subatomic Physics, Amsterdam, The Netherlands\\
% ...
$ ^{46}$H.H. Wills Physics Laboratory, University of Bristol, Bristol, United Kingdom\\
% ...
$ ^{51}$School of Physics and Astronomy, University of Glasgow, Glasgow, United Kingdom\\
$ ^{52}$Oliver Lodge Laboratory, University of Liverpool, Liverpool, United Kingdom\\
% ...
}
\end{flushleft}


\end{document}